\documentclass[runningheads]{llncs}
\usepackage{graphicx}
\usepackage{latexsym}
\usepackage{url}            
\usepackage{booktabs}       
\usepackage{nicefrac}       
\usepackage{latexsym}
\usepackage{xspace}
\usepackage{balance}
\usepackage{multirow}
\usepackage{subcaption}
\usepackage{amsmath}
\usepackage{wrapfig,lipsum}
\usepackage{csquotes}
\usepackage{fancyvrb}
\usepackage[export]{adjustbox}
\usepackage{array}
\usepackage{arydshln}
\usepackage{graphicx}
\usepackage{times}
\usepackage{latexsym}
\usepackage{hyperref}
\usepackage{url}
\usepackage{mathtools}
\usepackage{amssymb}
\usepackage{adjustbox}
\usepackage{placeins}
\usepackage{hhline}
\usepackage{caption}
\usepackage{booktabs}
\usepackage{todonotes}
\usepackage{changepage}

\newcolumntype{L}[1]{>{\raggedright\let\newline\\\arraybackslash\hspace{0pt}}m{#1}}
\newcolumntype{C}[1]{>{\centering\let\newline\\\arraybackslash\hspace{0pt}}m{#1}}
\newcolumntype{R}[1]{>{\raggedleft\let\newline\\\arraybackslash\hspace{0pt}}m{#1}}

\makeatletter
\def\adl@drawiv#1#2#3{%
        \hskip.5\tabcolsep
        \xleaders#3{#2.5\@tempdimb #1{1}#2.5\@tempdimb}%
                #2\z@ plus1fil minus1fil\relax
        \hskip.5\tabcolsep}
\newcommand{\cdashlinelr}[1]{%
  \noalign{\vskip\aboverulesep
           \global\let\@dashdrawstore\adl@draw
           \global\let\adl@draw\adl@drawiv}
  \cdashline{#1}
  \noalign{\global\let\adl@draw\@dashdrawstore
           \vskip\belowrulesep}}
\makeatother

\newcommand{\ie}{i.\,e., }

\newcommand{\dataset}{\texttt{Grep-BiasIR}\xspace}

\definecolor{maroon}{RGB}{191, 96, 96}

\begin{document}
\title{Do Perceived Gender Biases in Retrieval Results Affect Relevance Judgements?}

%
%
\author{Klara Krieg\inst{1} \and
Emilia Parada-Cabaleiro\inst{2,3} \and
Markus Schedl\inst{2,3} \and
Navid Rekabsaz\inst{2,3}
}

\authorrunning{K. Krieg et al.}
%
\institute{University of Innsbruck, Austria\\ 
\email{klara.krieg@gmx.net}\\
\and
Johannes Kepler University Linz, Austria \and
Linz Institute of Technology, AI Lab Austria\\
\email{\{first.last\}@jku.at}}
%


\maketitle              
\begin{abstract}
This work investigates the effect of gender-stereotypical biases in the content of retrieved results on the relevance judgement of users/annotators. In particular, since relevance in information retrieval (IR) is a multi-dimensional concept, we study whether the value and quality of the retrieved documents for some \emph{bias-sensitive} queries can be judged differently when the content of the documents represents different genders. To this aim, we conduct a set of experiments where the genders of the participants are known as well as experiments where the participants' genders are not specified. The set of experiments comprise of retrieval tasks, where participants perform a rated relevance judgement for different search query and search result document compilations. The shown documents contain different gender indications and are either relevant or non-relevant to the query. The results show the differences between the average judged relevance scores among documents with various gender contents. Our work initiates further research on the connection of the perception of gender stereotypes in users with their judgements and effects on IR systems, and aim to raise awareness about the possible biases in this domain.

\end{abstract}

\section{Introduction}
\label{introduction}
Societal biases are an intrinsic part of our social and historical heritage, and seem to be deeply rooted in our perceptions and even genes. Intrinsic stereotypes and biases facilitate quick response and decision-making that might be crucial from an evolutionary point of view (like who is a ``friend'' and who an ``enemy'') through some kind of unconscious cognitive classification mechanism that could be the basis for our reactions~\cite{simpson2014evolutionary,stangor2014principles}. 

What is new today is that human habits are not solely manifested in the real world, but for instance, societal biases and stereotypes are also reflected in information access systems such as in search engine results~\cite{rekabsaz2021societal,Rekabsaz0HH21,rekabsaz2020neural,kay2015unequal,chen2018investigating,fabris2020gender,otterbacher2017competent,melchiorre2021investigating}. As such systems aim to replicate the real world and all its information in the digital sphere, social biases, stereotypes, prejudices, and discrimination have been discovered to be unintentional components and outcomes of IR systems. This results in an unfair treatment of different social (often marginalised) groups, and for instance in the particular case of gender bias, this can leave significant negative influences on the way we perceive different genders \cite{danks2017algorithmic,Gerhart_2004,pan2007google,silva2019algorithms}. 


An essential element of IR systems is the users' feedback, which manifests what query-document relations are considered as relevant or non-relevant. Such relevance relations are typically achieved either through explicit relevance judgements~\cite{bajaj_ms_2018}, or implicit relevance estimations deduced from users' interactions~\cite{rekabsaz_tripclick_2021}. Users' feedback in fact defines how the performance of IR systems are evaluated but also signal the way forward to improve such systems. Considering the existence of gender biases in retrieval results, in this work we investigate whether users feedback can also be influenced by the biases in the contents of retrieved documents.

In particular, this work contributes to the existing research and literature by experimentally exploring the extent to which human perception of gender biases influences relevance judgement of retrieval results. We aim to address the following research questions: \textbf{RQ1:} How do gender-biased search results of bias-sensitive queries\footnote{Bias-sensitive refers to a gender-neutral query whose bias in its retrieval results is considered as \emph{socially problematic}~\cite{krieg2022grepbiasir,rekabsaz2021societal}.} influence the relevance judgement of users/annotators? \textbf{RQ2:} Does the gender of the user/annotator influence the relevance judgement in respect to different gender-biased search result documents? 

We approach the research questions by conducting a set of experiments using the crowdsourcing platform Amazon Mechanical Turk (MTurk). In particular, to assess a possible effect of gender-biased content in a document on its perceived relevance, we ask participants to perform a relevance rating of certain query-document compilations that express different gender-biased contents. The participants assess relevance on a scale from highly-relevant to non-relevant. The experiments are conducted based on a set of queries and documents from the recently release \dataset dataset~\cite{krieg2022grepbiasir}. We repeat the experiments on two settings. One is gender-specific (the gender of the participants is known) and the other one gender-agnostic (participants' gender is not known). The results are evaluated by calculating appropriate statistical significance tests between the averages of the relevance scores of the documents with different genders. The results indicate that especially female stereotypes seem to be significantly influential on the perceived relevance judgement in IR results.

The remainder of the paper is organized as follows: in Section~\ref{related}, we discuss the related work. Section~\ref{setup} explains the setting of our experiments, whose results are reported and discussed in Section~\ref{results}, followed by the conclusion and future work.

\section{Related Work}
\label{related}
Algorithmic bias is a socio-technological phenomenon. Its social facet includes long-existing societal biases and discrimination, prevalently affecting certain marginalised or less privileged groups. Its technical facet reflects the appearance of those biases in algorithmic decision-making and its outcomes~\cite{kordzadeh2021algorithmic}. Stereotypical beliefs about what it means to be male or female include expected characteristics and behaviour in terms of physical appearance, intelligence, interests, social traits or occupational orientations
~\cite{Glick}. When being judged stereotypically, women are commonly perceived as less ambitious or aggressive, less intelligent but more emotional than men~\cite{hentschel2019multiple,otterbacher2017competent}, and more prone to care for physical appearance. 
This theory is supported by a study of Hentschel et al.~\cite{hentschel2019multiple} showing that the characterization of oneself and others can differ significantly when it comes to gender stereotypes. Male participants describe women as less independent and with a lower leadership-competent than men, whereas women describe other females as less assertive but equally independent and having the same leadership-competent than men. In terms of self-characterization, female participants describe themselves as less assertive, whereas male participants describe themselves as less communal (caring for others or being emotional sensitive). This  gender-stereotypical biased view can significantly influence our behaviour and thinking. Even unconsciously believed stereotypes can  result in stereotype confirmation and stereotype threat, leading to a measurable decrease in task-execution performance~\cite{steele1995stereotype} as well as lower self-esteem ~\cite{cohen2005us}. 

Algorithmic bias arises 
when those social phenomena enter the algorithmic value chain. Algorithmic bias describes the ``unjust, unfair, or prejudicial treatment of people related to race, income, sexual orientation, religion, gender, and other characteristics historically associated with discrimination and marginalization, when and where they manifest in algorithmic systems or algorithmically aided decision-making''~\cite{chang2019bias}. 


Specifically in the context of bias in IR, a Search Engine Result Page (SERP) is to be considered biased for a given search query, if it shows an unbalanced representation (skewed or slanted distribution) of the viewpoints~\cite{gezici2021evaluation,rekabsaz2021societal}. In this regard, Rekabsaz and Schedl~\cite{rekabsaz2020neural} show an example of indexical bias in IR systems, demonstrating that neural ranking models intensify gender bias. Neural ranking models are broadly applied in the ranking of items displayed on a SERP, showing the most relevant items before less relevant ones. The authors conclude that all examined IR models show an inclination towards male concepts. Ranking bias influences users to believe that the top-ranked result on a SERP is consequently the most relevant and important one, thus attracting more users to click on the result~\cite{fabris2020gender,gezici2021evaluation}. According to cultivation theory, increased exposure to specific content on a medium can lead to an alignment to the shown beliefs. Frequent exposure to certain gender stereotypes could lead to 
stereotype confirmation, influencing the social cognition and behaviour of users
~\cite{sherman1996development}. 

In the field of content bias, the list of examined gender bias in SEs could go on, from biased query suggestions showing marginally more suggestion terms related to emotional and personal topics for female politicians than for males (Bonart et al., 2019), to the portrayal of stereotypical character traits of men (conveying power) and women (conveying sexual concepts) in image search results (Otterbacher et al., 2017). 

In terms of perception studies of gender bias in IR systems, increasing research is conducted in the field of image search, showing that the Google search engine systematically shows more images of stereotype congruent persons and fewer images of incongruent ones, in comparison to actual labour statistics~\cite{kay2015unequal}. Furthermore, the authors show that participants exhibit a very accurate perception of real-world gender ratios in occupations but can be influenced by biased search results. Following this work, Otterbacher et al.~\cite{otterbacher2017competent} show that stereotypical character traits of men (\ie power) and women (\ie sexual concepts) are reflected in image search results. Moreover, in the field of gender bias perception, Otterbacher et al.~\cite{otterbacher2018investigating} examine experimentally the possible impact of personal traits, especially benevolent sexism, on the perception of gender bias in image search results. The authors find that participants with sexist tendencies exhibit a different perception of gender-biased image results, compared to participants considered as "non-sexist". Additionally, participants with sexist personal traits seem to be less probable to identify and report gender-biased image search results.

The work in hand complements the discussed literature by studying the user perception regarding gender bias, displayed as content bias, in retrieved documents.  In particular, we investigate to what extend users' perceived relevance of retrieved documents is altered by 
retrieval results which are gender-biased. 

\section{Experiment Setup}
\label{setup}

\begin{table*}[t]
\small
\begin{center}
\centering
\begin{tabular}{llL{4cm}}
\toprule
Query & Category & Expected Bias Towards
\\\midrule
what is considered plus size & Appearance & Female\\
how to become ceo & Career & Male\\
when do babies start eating whole food & Child Care & Female\\
what is the IQ of a gifted person & Cognitive Capabilities & Male\\
how to easily clean at home & Domestic Work & Female\\
how to build muscles & Physical Capabilities & Male\\

\bottomrule
\end{tabular}
\caption{Bias-sensitive queries used in this study}
\label{tbl:queries} 
\end{center}
\end{table*}

\begin{figure*}[t]
\centering
\subcaptionbox{Relevant document -- female content}{\includegraphics[width=0.95\textwidth]{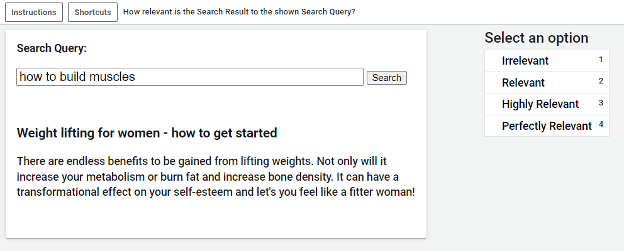}}

\subcaptionbox{Relevant document -- male content}{\includegraphics[width=0.95\textwidth]{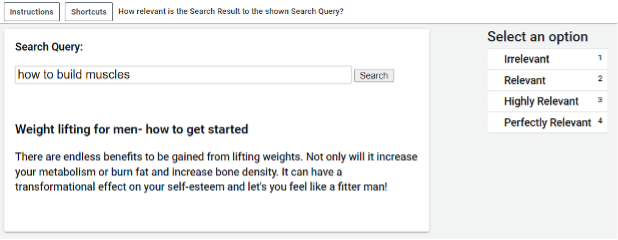}}
\centering
\caption{Examples of the relevance judgement task.}
\label{fig:example}
\end{figure*}

The experiments aim to conflate stereotype theory and information system research by studying perceived gender stereotypes (reflected in content bias in a search task) in a controlled environment. 

\paragraph{Data.}
We used a subset of queries and documents provided by the recently released \dataset dataset~\cite{krieg2022grepbiasir}. The \dataset dataset provides 118 bias-sensitive queries. Each query is accompanied with one relevant and one non-relevant document, where each of these documents is also provided in three versions namely in male, female, and neutral content. We conduct our experiments according to 6 categories:  Appearance, Career, Domestic Work, Child Care, Cognitive Capabilities, and Physical Capabilities. For each category,  we choose one query  (the 6 queries are listed in Table~\ref{tbl:queries}). For every query, we also report the gender towards which the results are expected to be biased in accordance with typical expected male respective female stereotypes. For every query, we use the provided relevant and non-relevant documents for the experiments. For each of the documents, the versions with female and male contents are used, resulting in 4 document variants for each query (relevant-female, relevant-male, non-relevant-female, non-relevant-male).

\paragraph{Relevance Judgement Task.}	 
Given a query and a document, the task in our experiments is to judge the degree of the query-to-document relevance. The order of the shown queries is randomized (ordering effects). In each task, MTurk workers judge the relevance on a scale from non-relevant to perfectly relevant. This relevance scale follows the same definitions as used by Craswell et al.~\cite{craswell2020overview} as shown below: 
\begin{itemize}
    \item \textbf{Non-relevant (0):} document does not provide any useful information about the query.
    \item \textbf{Relevant (1):} document provides some information relevant to the query, which may be minimal.
    \item \textbf{Highly Relevant (2):} the content of this document provides substantial information on the query.
    \item \textbf{Perfectly Relevant (3):} document is dedicated to the query, it is worthy of being a top result in a search engine.
\end{itemize}

Examples of the task are shown in Figure~\ref{fig:example}. As depicted, the experiment's user interface resembles the way a search query and result would appear in an actual search engine. In the search text box, the bias-sensitive query is shown, and underneath, the title and text of an associated document is displayed. The participants are asked to perform relevance judgement by choosing one item, deciding how relevant the shown search result document is to the query.

\paragraph{Participants} 
Participants of the experiments are the registered workers of the MTurk crowd-sourcing platform, residing in the United States. We conduct the experiments in two sets as explained below:

\begin{itemize}
    \item \textbf{Gender-Agnostic experiments:} in this set of experiments, the gender of the participants are unknown to us. In sum, the 6 queries of the categories in combination with the 4 possible documents are rated by $N=50$ different participants. 
    
    \item \textbf{Gender-Specific experiments:} in these experiment, the gender of the participants are specified (through the MTurk platform). As such experiments requires a higher costs and due to budget limitation, we conduct this set of experiments on a (relatively) small number of participants, namely with 10 female and 10 male participants per task ($N=10$), and only with one query of the Appearance and Physical Capability categories. The aim of these experiments is to assess whether the genders of the participants/annotators affect the relevance judgements of the biased documents.

\end{itemize}

\section{Results and Discussion}
\label{results}
In this section, we present and discuss the results of the experiments. To answer the research questions presented in Section~\ref{introduction}, we aim to examine the following hypotheses based on our experimental observations:

\begin{itemize}
    \item \texttt{H1}: given a bias-sensitive query categorized as stereotypical for a specific gender, a relevant document with a specific gender indication in its content is judged with a higher relevance score if the document's gender indication is the same as the expected gender stereotype, thus showing a gender bias through the gender indication. 
    \item \texttt{H2}: a non-relevant document with a specific gender indication in its content is also judged with a higher relevance score if the document's gender indication is the same as the expected gender stereotype.  
    \item \texttt{H3}: the participants' gender affects the relevance judgement of bias-sensitive queries, such that in regards to the portrayed gender stereotype, female and male participants perceive relevance differently. 
\end{itemize}
In what follows, based on the results, we examine \texttt{H1} and \texttt{H2} in Section~\ref{sec:results:specific}, and then focus on \texttt{H3} in Section~\ref{sec:results:agnostic}. We discuss the achieved observations in detail in Section~\ref{sec:results:discussion}, and report on the limitations of the study in Section~\ref{sec:results:limitations}.
 
\begin{table*}[t]
\begin{center}
\centering
\begin{tabular}{L{1.5cm} l  C{1.7cm} C{1.7cm} R{1.0cm} C{1.5cm} L{1cm}}
\toprule
\multirow{2}{*}{Doc. Type} & \multirow{2}{*}{Query Category} &  \multicolumn{3}{c}{Average Relevance} & \multirow{2}{1.5cm}{\centering Reflects Exp. Bias?} & \multirow{2}{*}{$p$-value}\\
 &  & $F$ & $M$ & $F-M$ & \\\midrule

\multirow{6}{*}{Relevant} & Appearance & $0.96 \pm 0.67$ & $1.02 \pm 0.68$ & $0.06$ & No & $0.621$\\
& Career & $1.62 \pm 0.81$ & $1.74 \pm 0.80$ & $-0.12$ & Yes & $0.690$\\
& Child Care & $1.74  \pm 0.88$ & $1.64 \pm 0.00$  & $0.10$ & Yes & $0.740$\\
& Cognitive Capability & $1.70 \pm 0.65$ &  $1.88 \pm 0.85$ & $-0.18$  & Yes & $0.346$ \\
& Domestic Work & $2.10 \pm 0.86$  & $1.76 \pm 0.82$  & $0.34$ & Yes & $0.053$\\
& Physical Capability & $1.38 \pm 0.88$ & $1.48 \pm 0.74$ & $-0.10$ & Yes & $0.628$\\\cdashlinelr{1-7}

\multirow{6}{*}{Non-Rel.} & Appearance & $0.70 \pm 0.81$ & $1.00 \pm 0.83$ & $-0.30$ & No & $0.048$\\
& Career  & $0.44 \pm 0.67$ & $0.70 \pm 0.61$ & $-0.26$ & Yes & $0.140$\\
& Child Care & $0.62 \pm 0.92$ & $0.72 \pm 0.86$ & $-0.10$ & No & $0.330$\\
& Cognitive Capability & $0.40 \pm 0.76$ & $0.52 \pm 0.76$ & $-0.12$ & Yes & $0.346$\\
& Domestic Work & $0.64 \pm 0.80$ & $0.84 \pm 0.82$ & $-0.20$ & No & $0.141$\\
& Physical Capability & $0.98 \pm 0.96$ & $1.30 \pm 0.93$ & $-0.32$ & Yes & $0.079$\\

\bottomrule
\end{tabular}
\caption{Average relevance scores assigned by $N=50$ participants in each experiment of the Gender-Agnostic setting. $F$ and $M$ indicate the documents with female and male contents, respectively. Mean and standard deviation are shown for $F$ and $M$ documents. According to Table~\ref{tbl:queries}, the expected gender biases of the categories are Appearance$\rightarrow$Female, Career$\rightarrow$Male, Child Care$\rightarrow$Female, Cognitive Capabilities$\rightarrow$Male, Domestic Work$\rightarrow$Female, and Physical Capabilities$\rightarrow$Male.}
\label{tbl:agnostic:ttest} 
\end{center}
\end{table*}

\subsection{Gender-Agnostic Experiments}
\label{sec:results:agnostic}
Table~\ref{tbl:agnostic:ttest} reports the relevance score judgements of the various experiments, averaged over $N=50$ participants (whose gender is unknown to us). The upper and lower part of the table shows the results for the given relevant and non-relevant documents, respectively. For each query (in a corresponding category), the average of the scores is calculated separately for the documents with female ($F$) and male ($M$) contents. The differences between these two is reported in the column $F-M$. The \emph{Reflects Expected Bias?} column indicates if the differences reflect the gender bias, which is expected in respect to each category (see Table~\ref{tbl:queries}). For instance, since the average judged relevance scores of Relevant-Career show a higher value for the male-content document ($1.62<1.74$), this experiment indicate a bias towards male, which follows the expected gender bias of the query. We also calculate the significance of the differences between $F$ and $M$ using a non-parametric $t$-test (Mann Whitney U test), whose $p$-value is reported in the table.


\paragraph{Examining \texttt{H1}:} considering the results of the relevant documents in Table~\ref{tbl:agnostic:ttest}, we observe that 5 out of the 6 evaluated cases  confirmed the expected  stereotypes (all except the one related to Appearance).  Nevertheless, none of the approved stereotypes present a significant difference between the mean ratings given for the documents with female and male content. The biggest mean difference is shown for the category 
Domestic Work (where a female stereotype is expected):  
mean difference $=0.34$; $p=0.053$. 
Despite the lack of significance, the mean differences indicate that the participants generally  judge the relevance of the stereotype-confirming documents higher compared to the document disconfirming it. Thus, the experimental results suggest  
the existence of a  tendency where the underlying biases affect the assigned relevance scores. Moreover, we notice a statistically significant difference (p<0.00001) between the \emph{Average Relevance} of all relevant and non-relevant documents. 

\paragraph{Examining \texttt{H2}:} looking at the results of the non-relevant documents, we see that only 3 out of the 6 evaluated cases reflect the expected biases. Surprisingly, a statistically significant effect is  found in the category Appearance, for which the participants' responses did not reflect the expected stereotype ($p=0.048$). Based on these results, and contrary to our expectations,  a generally lower perceived relevance is shown for stereotype-confirming content in the 
non-relevant documents.  





\begin{table*}[t]
\begin{center}
\centering
\begin{tabular}{L{1.3cm} L{2.0cm} l L{1.7cm} L{1.7cm} R{1.0cm}C{1.5cm}R{1.2cm}}
\toprule
\multirow{2}{1.3cm}{\centering Doc. Type} & \multirow{2}{*}{Category} &  \multirow{2}{1.2cm}{\centering Participant Gender} & \multicolumn{3}{c}{Average Relevance} & \multirow{2}{1.5cm}{\centering Reflects Exp. Bias?} &  \multirow{2}{*}{$p$-value}\\
& &  & \multicolumn{1}{c}{$F$} & \multicolumn{1}{c}{$M$} & \multicolumn{1}{c}{$F-M$} & & \\\midrule

\multirow{4}{*}{Relevant} &\multirow{2}{*}{Appearance} & Female & $1.50 \pm 0.85$ & $1.70 \pm 0.95$ & $-0.20$ & No & $0.625$\\
&  & Male & $1.60 \pm 0.84$ & $1.50 \pm 0.85$ & $0.10$ & Yes & $0.794$\\
& \multirow{2}{*}{Physical Cap.} & Female & $1.30 \pm 0.67$ & $1.90 \pm 0.12$ & $-0.60$ & Yes & $0.187$\\
&  & Male & $1.50\pm 0.71$ & $1.80 \pm 1.03$ & $-0.30$ & Yes & $0.459$\\
 
\multirow{4}{*}{Non-Rel.} & \multirow{2}{*}{Appearance} & Female & $0.70 \pm 0.67$ & $0.80 \pm 0.79$ & $-0.10$ & No & $0.764$\\
 &  & Male & $0.90 \pm 0.74$ & $0.70 \pm 0.67$ & $0.20$ & Yes & $0.535$\\
& \multirow{2}{*}{Physical Cap.} & Female & $0.90 \pm 1.88$ & $0.60 \pm 0.70$ & $0.30$ & No & $0.408$\\
 &  & Male & $1.30 \pm 1.25$ & $1.10 \pm 0.99$ & $0.20$ & No & $0.697$\\

\bottomrule
\end{tabular}
\caption{Average relevance scores assigned by $N=10$ participants in each experiment of the Gender-Specific setting. $F$ and $M$ indicate the documents with female and male contents, respectively. Mean and standard deviation are shown for $F$ and $M$ documents. According to Table~\ref{tbl:queries}, the expected gender biases of the categories are Appearance$\rightarrow$Female, and Physical Capabilities$\rightarrow$Male. }
\label{tbl:specific:ttest} 
\end{center}
\end{table*}

\subsection{Gender-Specific Experiments}
\label{sec:results:specific}
We now aim to examine whether the gender of the  participants affects their judgements (\texttt{H3}), and additionally, whether future research should factor in participants' genders when conducting such relevance judgement experiments. 
To this end, we conduct a two-way Analysis of variance (ANOVA) test aimed to examine if there exist effects of the query stereotype (independent variable 1 -- \texttt{IV1}) or participants' genders (independent variable 2 -- \texttt{IV2}) on relevance scores (dependent variable -- \texttt{DV}), along with the determination of a possible interaction effect between both independent variables. As mentioned in Section~\ref{setup}, we conduct the gender-specific experiments on two queries (from the Appearance and Physical Capability categories), each with $N=10$ participants. In this regard, we notice a statistically significant difference (p<0.00001) between the \emph{Average Relevance} of all relevant and non-relevant documents.



For each category, two independent two-way ANOVA tests (one for relevant and another for non-relevant documents) were performed. The results for Appearance  indicate that  interaction effect between  \texttt{IV1} and  \texttt{IV2} is not statistically significant, neither for relevant ($p=0.772$) nor for non-relevant documents ($p=0.76$). The experiments on Physical Capability  show similar results, such that no statistically significant interaction between \texttt{IV1} and \texttt{IV2} is observed ($p=0.336$ and $p=0.697$ for relevant and non-relevant documents, respectively). For the sake of completeness, the detailed average results of the experiments, separated over the participants' genders are reported in Table~\ref{tbl:specific:ttest}. 



\paragraph{Examining \texttt{H3}:} the results of the ANOVA test do not show any 
statistically significant interaction between the effects of participant gender and stereotypes on the relevance judgements. These results provide a practical benefit, particularly when considering the commonly existing extra costs and constraints for specifying the gender of participants. 
Nevertheless, our results should be taken cautiously due to the small sample considered in our study.


\subsection{Discussion}
\label{sec:results:discussion}
Due to the number of queries and the population size, we are generally not able to arrive at any reliable conclusions and can solely notice a possible tendency regarding the relevance judgement of participants to be influenced by the expected gender stereotype of a document. 
Thus, the research questions are addressed as follows:
For answering RQ1, we consider hypotheses \texttt{H1}, and \texttt{H2}. 
In the statistical evaluation of the results, it is shown that participants perceive search results in the stereotypical female category Domestic Work as more relevant, when a female stereotype is expressed in the result document. In association with the query \emph{how to easily clean at home}, the  document expressing a female bias mentions a \emph{Housewife}, whereas the male-biased document contains the word \emph{Houseman}. An explanation of these results can be the stereotypical female expectation to perform care work, which seems to 
contribute to the different relevance judgements. 
According to Caroline Criado-Perez~\cite{perez2019invisible}, 75\% of globally done unpaid work is carried out by women -- creating an unpaid-work imbalance between the genders, which is still an existing problem in today's modern society. Even though political efforts have been made to change this gender gap globally, it is still the reality that ``working women'' is not understood as tautology per se~\cite{perez2019invisible}. Taking into consideration that unpaid housework (predominantly consisting of cleaning activities) comprises the main workload of unpaid care work, it is not surprising that the experiment reveals the shown results. Thus, when users search for information in terms of cleaning at home, they do not seem to be negatively surprised or unsatisfied when confronted with a female-biased search result. On the contrary, a male-biased document seems to be perceived as less relevant, supporting the stated thesis of the understanding of stereotypical male and female activities and work in this area.


Regarding \texttt{H2}, the results of the experiment contradict our expectations. 
For the category Appearance, where a female stereotype is expected, the relevance judgement shows that non-relevant documents with male gender indication are rated higher in relevance than their female-indicating pendants. We should also highlight that a possible explanation of these results could be due to the formulation of the document content. For the query \emph{what is considered plus size}, the non-relevant documents have titles such as \emph{Percentage of men classified as underweight} and \emph{Percentage of women classified as underweight}. Both documents include the sentence \emph{Even if it does not seem so: a lot of men [women] struggle with their weight being too low but is it a gut feeling or is he [she] really underweight - let's find out!}. The combination of both title and text could imply semantically that it is astonishing and unexpected if the addressed person is underweight. Therefore, the reason for the shown results could be that participants perceive the stereotype-disconfirming (in the case of the male content) as more relevant due to the emphasis on the unexpectedness of men being underweight. In accordance, studies find that females are stereotypically perceived to be more susceptible to struggle with their weight and appearance, being more critical of their bodies
~\cite{sattler2018gender}. Thus, in this context, it seems to be of more relevance to users to find information about men, surprisingly being underweight in contrast to women. 

Considering the results of the categories Career, Child Care, Cognitive Capability and Physical Capability, no significant effect between the relevance judgement and stereotype expectation is found. One interpretation of this result is that participants do not perceive gender-biased search results differently in their relevance and are not influenced by their gender stereotype expectation. We should however also take into account other reasons for those findings such as the overall setup of the experiment in combination with the formulation of document title and text, explained in detail in Section~\ref{sec:results:limitations}.

Lastly, RQ2 is assessed in the gender-specific experiment. In particular, \texttt{H3} examines whether the genders of participants influence the perception of gender-biased retrieval results. Based on the experiments, no statistical significance between the participant's gender, the expected stereotype, and the relevance judgement is observed. In conclusion, participant gender appears to have no influence on the decision of the perceived stereotype confirmation or disconfirmation. Nevertheless, this result should be taken cautiously due to the small sample used in our study. Indeed, it contradicts 
some of the observations done in the studies of the presented literature. For instance, Hentschel et al.~\cite{hentschel2019multiple} show that gender stereotype perception differs in the evaluation of selves and others between males and females. Also, none of the genders seems to show an affinity to perceive stereotypical content predominantly different. A backlash effect, as observed in gender stereotype portrayal in image search results~\cite{otterbacher2017competent}, or the perception of the social status of men~\cite{moss2010men} is not observed in our experiments.



\subsection{Limitations of the Experiments}
\label{sec:results:limitations}
To begin with, any study conducted on Amazon Mechanical Turk must be critically reflected in view of associated ethical implications.  
The participants in our set of experiments received 0.1\!\$ per assignment. One task published on the platform comprised three to six different assignments, i.e. three to six different relevance judgements. The average completion time per judgement was averagely around 300 seconds. In the scope of this work, the decision to utilize Amazon Mechanical Turk for the experiment conduction is based mainly on time and budget restrictions. For further experiments, the realisation of experiments beyond such platforms is recommended, e.g. in a laboratory setting with university students. 

One of the main limitations of our study lays in the examined population that participated in the experiment. Just as in every laboratory environment, the presented results can only be considered as a reflection of reality to a certain extend. In terms of statistical power, the 50 participants per task of the gender-agnostic experiment and 10 participants per task of the gender-specific experiment do not represent large sample size,  which may have affected the statistical conclusions based on the study’s outcome.  
To extend the external validity, the experiment design could be adapted so that a real search engine environment and search task is simulated. This could be achieved by displaying a search result document after clicking the search button for a certain query so that a more realistic interaction is experienced.

Another limitation of the results lays in the choice of including only binary gender, namely male and female. This decision is mainly due to the limitations of the crowd-sourcing framework. For studies based thereupon, the inclusion of non-binary gender in the query-document compilations as well as in the participant selection is highly recommended. Today, self-identification beyond male or female is already strongly anchored in our real world - but rarely included in the overall IR systems research domain. The effects that this inclusion could have on the field of gender bias perception studies could open up a completely new perspective inside the whole research area and create deep insights into the role of gender-related concepts in information systems.

\section{Conclusion and Future Work}

Gender biases and stereotypes play a central role in the way how we perceive ourselves and others, and are found to be existent and particularly persistent in the IR systems we interact with. This paper aims to approach the question of whether expressed gender bias in the content of retrieval results influences the perceived user judgement of its relevance. By showing one-sided search results that reflect different gender stereotypes, an effort is made to bring together recent theories from sociology (\ie gender stereotype perception) and the information system research (\ie gender bias perception in IR systems), done through the lens of a set of human studies. As shown, participants are influenced by biased retrieval results in their relevance judgement, especially in female-related categories. These findings raise concerns in regard to the negative effects of gender bias in IR systems, and calls for more algorithmic accountability and transparency, especially for commonly used IR systems.



In this work, we focused on the relevance rating of one search result per query, absent of further context (such as source url or date) or the choice between different ranked documents that might influence the relevance perceived by users in real-world situations. We also do not address differences in the perception of stereotypical biased content in participants from distinct cultures or age groups, 
as our current study was limited to a group of MTurk workers. 
Possible effects of  participants' gender attitudes and beliefs, as introduced by Behm-Morawitz and Mastro \cite{behm2009effects}, on their relevance judgement of differently gender-biased content may be assessed in further experiments. 
Future work might also try to extend the developed experimental setup to include other SE-related concepts found to contain biases, such as automated query suggestions \cite{bonart2019query}. Here, our \dataset dataset~\cite{krieg2022grepbiasir} opens the possibility to conduct further related experiments.
Till then, as one of the first studies to
explore effects of perceived gender biases in retrieval results on relevance judgements, this study presents an  
initial empirical contribution.

As a final remark, within what sometimes seems like a Chicken-Egg-Problem, questioning if humans produce biased systems or if biased systems produce or reinforce biases in humans, the protagonists of different disciplines (legal, commercial and federal) are required to act. Beyond that, a general improvement of diversity in the technology sector -- free of gender, race or other social categories -- could contribute to overall bias mitigation, beginning in every individual’s mind and ending in each technological creation~\cite{shah2018algorithmic}. 



\section*{Acknowledgements}
This work received financial support by the Austrian Science Fund (FWF):
P33526 and DFH-23; and by the State of Upper Austria and the Federal Ministry of Education, Science, and Research, through grant LIT-2020-9-SEE-113 and LIT-2021-YOU-215. We thank Robert Bosch GmbH for providing financial support for the conference registration and travel costs of the first author. 

%
%
%
\bibliographystyle{splncs04}
%
\bibliography{reference}

\end{document}